# Are high-energy photoemission final states free-electron-like?

V.N. Strocov,[1] L.L. Lev,[1,2] F. Alarab,[1] P. Constantinou,[1] T. Schmitt,[1]
T. J. Z. Stock,[3] L. Nicolaï,[4] J. Očenášek[4] & J. Minár[4]

[1]Swiss Light Source, Paul Scherrer Institute, CH-5232 Villigen-PSI, Switzerland
[2]Moscow Institute of Physics and Technology, Dolgoprudny, Moscow Region 141701, Russia
[3]London Centre for Nanotechnology, University College London, London WC1H 0AH, UK
[4]University of West Bohemia, New Technologies Research Centre, 301 00 Plzeň, Czech Republic

## Abstract

Three-dimensional (3D) electronic band structure is fundamental for understanding a vast diversity of physical phenomena in solid-state systems, including topological phases, interlayer interactions in van der Waals materials, dimensionality-driven phase transitions, etc. Interpretation of ARPES data in terms of 3D electron dispersions is commonly based on the free-electron approximation for the photoemission final states. Our soft-X-ray ARPES data on Ag metal reveals, however, that even at high excitation energies the final states can be a way more complex, incorporating several Bloch waves with different out-of-plane momenta. Such multiband final states manifest themselves as a complex structure and excessive broadening of the spectral peaks from 3D electron states. We analyse the origins of this phenomenon, and trace it to other materials such as Si and GaN. Our findings are essential for accurate determination of the 3D band structure over a wide range of materials and excitation energies in the ARPES experiment.

# Introduction

Knowledge of electronic band structure resolved in three-dimensional (3D) electron momentum (**k**) is fundamental for understanding a vast diversity of physical phenomena in crystalline solid-state systems. Recently, the interest in 3D band structure has been boosted due to its essential role in topological phases such as Weyl semimetals characterised by 3D cones of linear electron dispersion (see, for example, refs. 1,2) as well as their generalisation to high-fold chiral fermions[3,4] and high-dimensional degeneracies such as the Hopf links and nodal lines, chains and knots in 3D **k**-space (see the reviews[5–8] and the references therein). Less straightforward but equally important implications of the 3D band structure include, for example, interlayer interaction and 3D charge-density waves in van der Waals materials[9–11], formation of quantum-well states at interfaces and heterostructures[12–16] as well as minibands in semiconductor superlattices[17], **k**-dependent electron-phonon interactions[18], dimensionality-driven phase transitions[19,20], 3D quantum Hall effect[21], and many more properties of solid-state systems.

High-energy angle-resolved photoelectron spectroscopy (ARPES), operating in the soft- and hard-X-ray photon energy ($hv$) regions, has pushed the **k**-resolving spectroscopic abilities of this technique from the conventional surface science to the intrinsic electronic structure deep in the bulk, buried interfaces and heterostructures, and diluted impurity systems (see the recent reviews[22–26] and the references therein). The main advantage of high photoelectron energies is an increase of the photoelectron mean free path ($\lambda_{PE}$) to a few nanometres and more[27]. Crucial for the experimental determination of 3D band structure, the increase of $\lambda_{PE}$ translates, via the Heisenberg uncertainty principle, to sharpening of the intrinsic resolution of the ARPES experiment in the out-of-plane momentum ($k_z$) which is defined as $\Delta k_z = \lambda_{PE}^{-1}$ [28]. The sharp resolution in $k_z$ underlies the applications of high-energy ARPES for accurate determination of the electronic band structure resolved in 3D **k**-space as illustrated by many of the works cited above.

In contrast to the in-plane momentum $\mathbf{k}_{//} = (k_x, k_y)$, conserved in the photoemission process because of the in-plane periodicity of the system, the $k_z$ component is distorted upon the photoelectron escape from the crystal to vacuum. It can however be reconstructed based on its conservation in the photoexcitation process in the bulk (corrected for the photon momentum $\mathbf{p}_{hv}$) if the final-state $k_z$ is known. Conventionally, the final-sate dispersion is modelled within the free-electron (FE) approximation, where $k_z$ is found as $k_z = \frac{\sqrt{2m}}{\hbar}\sqrt{E_k - \frac{\hbar^2}{2m}K_{//}^2 - V_0}$, with $E_k$ and $\mathbf{K}_{//}$ being the photoelectron kinetic energy and in-plane momentum, respectively, $m$ the free-electron mass, and $V_0$ the inner potential. Somewhat stretching this formula, an energy dependence of the dynamic exchange-correlation[29,30] can be accommodated via an energy-dependent $V_0$. Importantly, the FE approximation implies that the final-state wavefunction is a plane wave, where the finite $\lambda_{PE}$ is described by an imaginary part of $k_z$. It has since long been realised that at low excitation energies used in the conventional VUV-ARPES the FE approximation may in many cases fail even for metals[31–34] and all the more for semiconductors[35] and more complex materials, for example, transition metal dichalcogenides[36–38]. For high-energy ARPES, however, the relevance of this approximation is commonly taken for granted. Being quintessential for 3D band mapping with high-energy ARPES, this assumption is based on a physically appealing argument that at high excitation energies $E_k$ of photoelectrons much exceeds modulations of the crystal potential $V(\mathbf{r})$, and they can be considered as free particles.

Here, we analyse soft-X-ray ARPES data on Ag the metal and demonstrate that even at high excitation energies the complexity of the final states can go far beyond the FE picture. In particular, they can be composed of multiple Bloch waves having different $k_z$s which manifest themselves as complex structure of the spectral peaks or their excessive broadening. This analysis extends to GaN and Si the semiconductors. We theoretically demonstrate the origin of these non-trivial effects as resulting from hybridization of plane waves on the crystal potential, and elucidate how they should be taken into account for accurate determination of 3D valence-band dispersions in the high-energy ARPES experiment.

# Results

Fig. 1 presents the Brillouin zone (BZ) of the fcc Ag (a) and the experimental out-of-plane cross-section of the Fermi surface (FS) in the ΓXW symmetry plane measured under variation of $hv$ (b). The indicated $k_z$s, running through a sequence of the Γ and X points, were rendered from the $hv$ values assuming FE final states with $V_0$ = 10 eV. In-plane cross-sections measured at two $hv$ values, bringing $k_z$ to the Γ and X points, are presented in the two panels (c). In general, the experimental out-of-plane FS follows a pattern of repeating rounded contours characteristic of the states near the Fermi level ($E_F$) formed by the sp-band of Ag. This pattern is reproduced by our one-step ARPES calculations (e) where FE-like final states were used. Surprisingly, however, a closer look at the experimental FS reveals significant deviations: (1) Multiple FS contours, offset in $k_z$, can be resolved in some ($E$,**k**) regions such as those marked by magenta arrows. The corresponding multiple dispersions coming from the sp-band are apparent, for example, in the ARPES image measured at $hv$ = 997 eV (d, top) and the corresponding momentum-distribution curve as a function of $k_x$ at $E_F$ ($k_x$-MDC, yellow line). This multiple-dispersion pattern contrasts to the clean dispersions at $hv$ = 894 eV (d, bottom). As we discuss in more detail below, such replica spectral structures demonstrate that the final states incorporate multiple bands with different $k_z$s – hereinafter called multiband final states (MBFSs) – which is a phenomenon beyond the conventional picture of FE-like final states implying one single band with one $k_z$. In our case the separation of the $k_z$s in these MBFSs is larger than the intrinsic $\Delta k_z$ (according to the $\lambda_{PE}$ values from the TPP-2M formula, varying from ~0.15 Å$^{-1}$ at 300 eV to 0.056 Å$^{-1}$ at 1300 eV); (2) The second type of deviations from the FE final states, seen in the out-of-plane FS (b), is a notable spectral intensity spreading into the X points where the sp-band is unoccupied. Furthermore, broadening of the FS contours in $k_z$ irregularly varies through **k**-space, and in some ($E$,**k**) regions (such as those marked by yellow arrows) can be excessively large. These two effects are also caused by the MBFSs, but in this case the $k_z$s are separated less than $\Delta k_z$. We note that in the extremes of the $E(k_z)$ dispersion (d$k_x$/d$k_z$=0 in the out-of-plane FS) the MBFSs have only a second-order effect on the ARPES structure; however, even in this situation a large enough $k_z$ separation within the MBFSs can cause multiple FS contours, as seen in the in-plane FS map measured at $hv$ = 712 eV (c, magenta arrow). Obviously, the MBFS effects are not reproduced by the ARPES calculations (e) employing FE final states. Although presently on a qualitative level, these effects are reproduced by our one-step ARPES calculations (Supplemental Material) where the final states are treated within the multiple-scattering formalism, naturally incorporating the non-FE effects including the MBFSs.

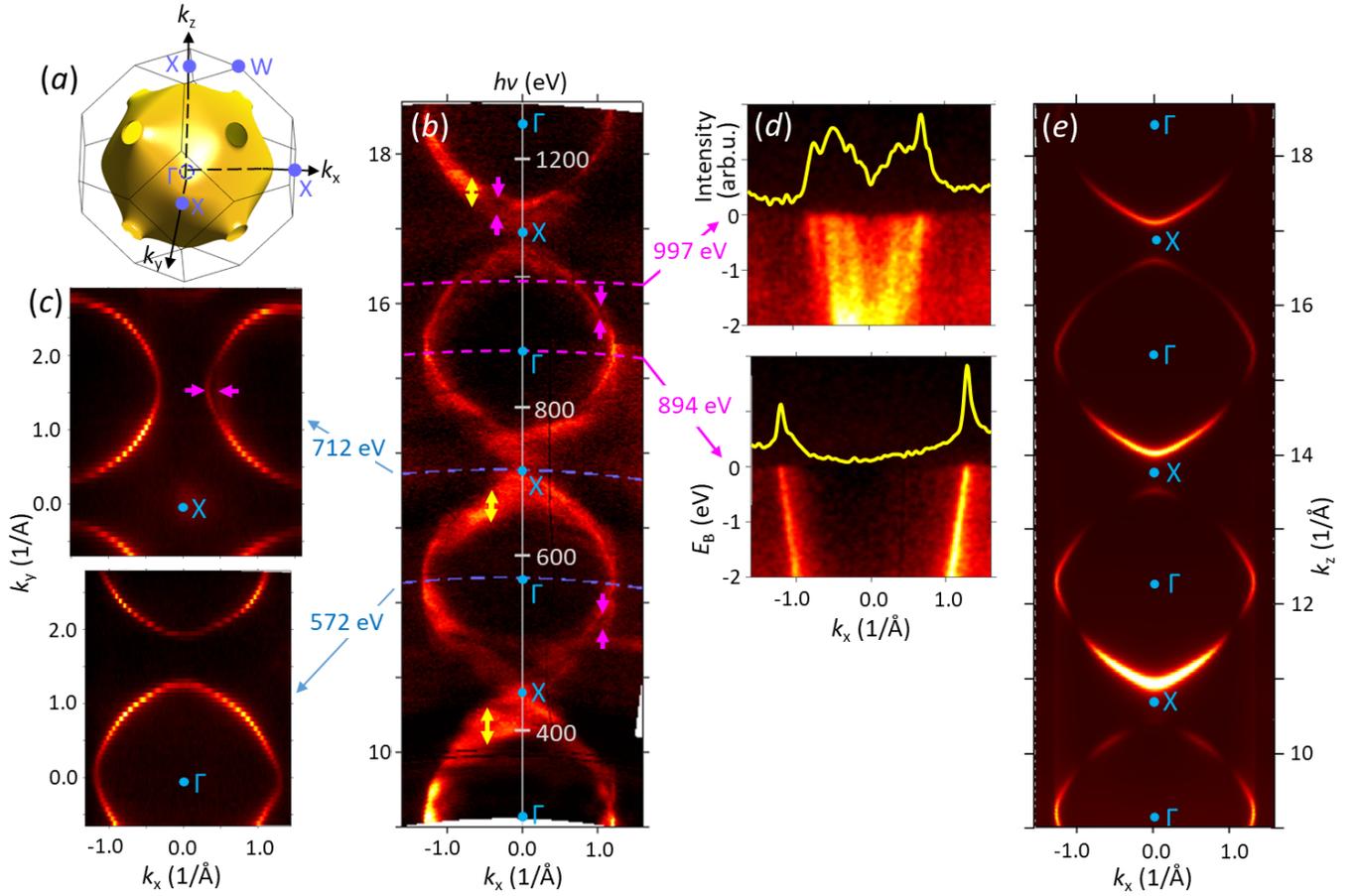

Fig. 1. FS cross-sections for Ag(100): Theoretical FS (a), its experimental out-of-plane cross-section (b), and two in-plane cross-sections (c) measured at the indicated $h\nu$ values, bringing $k_z$ to the Γ and X points (lower and upper panels, respectively). Replicas and broadening of the FS contours in certain ($E$,**k**) regions (such as those marked by magenta and yellow arrows, respectively) manifest MBFSs. These effects are particularly clear in the ARPES image and $k_x$-MDC at $h\nu$ = 997 eV (d, top) in contrast to those at $h\nu$ = 894 eV (bottom). These effects are beyond the one-step ARPES calculations with FE-like final states (e).

In Fig. 2, the theoretical $E(\mathbf{k})$ along the ΓX direction (a) is compared with the experimental out-of-plane band dispersions $E(k_z)$ at $k_x$=0 (b) and the in-plane $E(\mathbf{k}_{//})$ images (c) measured at $k_z$ running through the successive Γ points (energies as binding energies $E_b$ relative to $E_F$). Again, the gross structures of the experimental $E(k_z)$ follow the expected periodic pattern with the sp-band crossing $E_F$ as reproduced by our one-step ARPES calculations in (e) with the FE-like final states. We see, however, replicas and anomalous broadening of the sp-band (such as marked by magenta arrows) as well as significant spectral intensity around the X point. These anomalies appear most clearly in the zoom-in of the sp-band and the $k_z$-MDC at $E_F$ (d, yellow line) where we observe a complex multi-peak structure of the spectral intensity around the X point. Again, these effects are manifestations of the MBFSs, with the ARPES dispersions originating from the individual final-state bands marked by the magenta arrows. Again, they are absent in the ARPES calculations employing FE final states (e) but are qualitatively reproduced upon inclusion of multiple-scattering final states (Supplemental Material). The MBFS effects could not be observed in the first soft-X-ray study on Ag(100) focused on the 3d states[39] because the smaller $k_z$ dispersion of these states compared to the sp ones could not provide sufficient separation of the spectral peaks from the different bands in the MBFS. We note in passing that the experimental 3d states appear in ~1 eV below the LDA-DFT energies; such an energy shift, already noticed for Cu, is a pronounced self-energy effect due to non-local exchange interaction of the 3d electrons strongly localized in the core region[40].

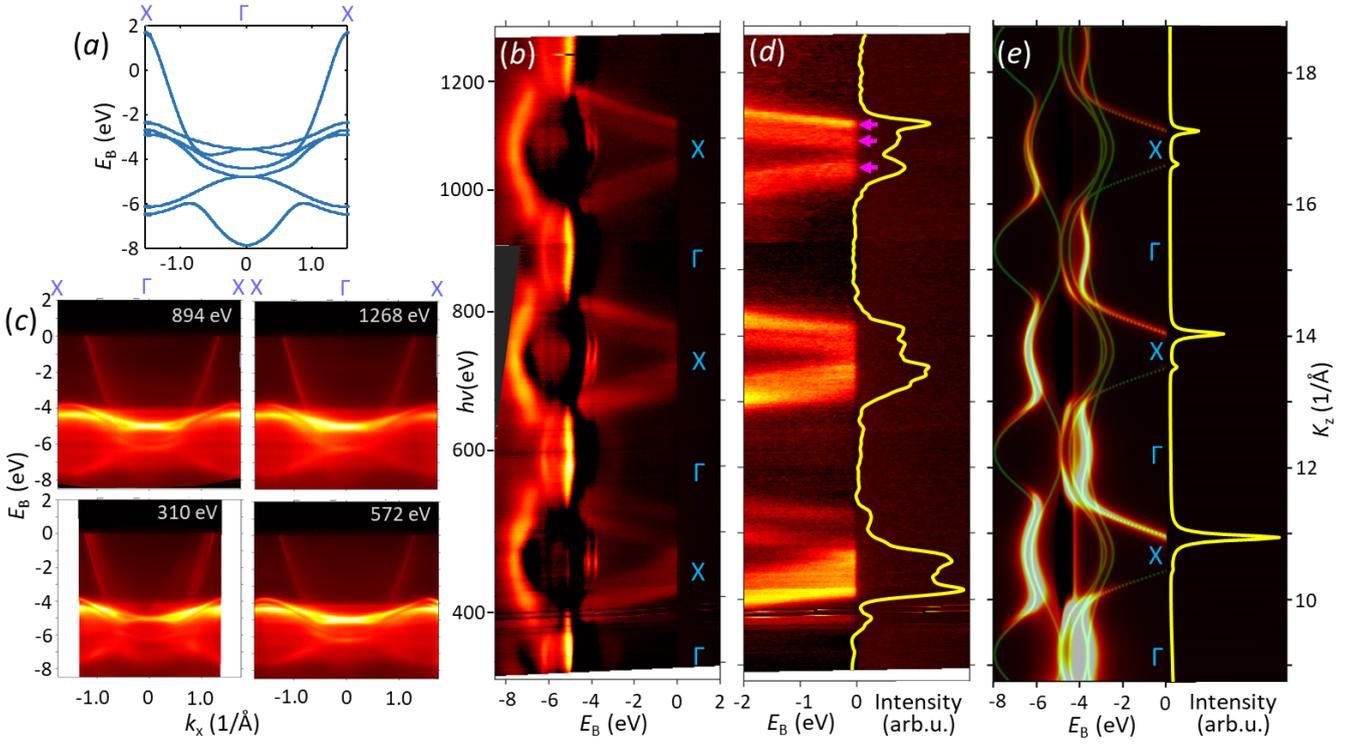

Fig. 2. Band dispersions along the ΓX direction for Ag(100): Theoretical $E(\mathbf{k})$ (a) compared with the experimental out-of-plane ARPES dispersions at $k_x=0$ (b, the spectral intensity represented in logarithmic scale) and (c) in-plane dispersions for the indicated $h\nu$ values, bringing $k_z$ to the successive Γ point. A zoom-in of the $sp$-band (d) shows its replicas and excessive broadening (such as marked by magenta arrows) most evident in the $k_z$-MDC at $E_F$ (yellow line) as multiple and broadened spectral peaks, manifesting the MBFSs. These effects are beyond the one-step calculations of the ARPES intensity and $k_z$-MDC with FE-like final states (e).

# Discussion

## Origin of the MBFSs

By definition, a FE-like final state in the crystal is one single plane wave $e^{i(\mathbf{k}+\mathbf{G})\mathbf{r}}$ which matches the outgoing photoelectron plane wave. In the whole multitude of bands, formally available under $E_k$ and $\mathbf{K}_{//}$ conservation, this plane wave corresponds to one single band that we will refer to as primary, relaying Mahan's primary photoemission cones [41]. All other bands in the multitude give strictly zero contribution to the photocurrent. We will be calling them secondary, relaying Mahan's secondary cones. The MBFS effects, observed in our ARPES data, indicate that the corresponding final states may include, for given $E_k$ and $\mathbf{K}_{//}$, several bands with different $k_z$s giving comparable contributions to the ARPES intensity. These effects obviously fall beyond the FE-like picture. As the first-principles calculations can not yet exhaustively describe our experimental results, we will analyse the MBFS effects based on insightful model calculations.

The non-FE effects in the final states, in particular their multiband composition, is certainly a phenomenon not new for low-energy ARPES. They have been studied experimentally and theoretically for 3D bulk band dispersions in various materials including Cu[31,32], Mg[34] and even Al the paradigm FE metal[14,42], semiconductors[35], various transition metal dichalcogenides[36–38] as well as surface states, in particular for the Al(100) and (111) surfaces[33]. However, it is intriguing to observe such effects in our soft-X-ray energy range. Why do they appear in spite of the fact that the photoelectron $E_k$ is overwhelmingly large compared to the $V(\mathbf{r})$ modulations?

We will now build a physically appealing picture of the non-FE effects in the photoemission final states using their standard treatment as the time-reversed LEED states[43]. They are superpositions of damped Bloch waves $\phi_\mathbf{k}(\mathbf{r})$ with complex $k_z$, whose imaginary part $\mathrm{Im}k_z$ represents the (1) inelastic electron scattering, described by a constant optical potential $V_i$ (imaginary part of the self-energy), and (2) elastic scattering off the crystal potential[44–47]. The amplitudes $A_\mathbf{k}$ of these $\phi_\mathbf{k}(\mathbf{r})$, determining their contribution to the total ARPES signal, were determined within the matching approach of the dynamic theory of LEED[17,38,44,45,48,49] where the electron wavefunction in the vacuum half-space (superposition of the incident plane wave $e^{i\mathbf{K}_0\mathbf{r}}$ and all diffracted ones $e^{i(\mathbf{K}+\mathbf{g})\mathbf{r}}$, $\mathbf{g}$ being the surface reciprocal vectors) is matched, at the crystal surface, to that in the crystal half-space (superposition of $\phi_\mathbf{k}(\mathbf{r})$ satisfying the surface-parallel momentum conservation $\mathbf{k}_{//}=\mathbf{K}_{//}+\mathbf{g}$). The underlying complex bandstructure calculations utilised the empirical-pseudopotential scheme, where $\phi_\mathbf{k}(\mathbf{r})$ are formed by hybridization of plane waves $e^{i(\mathbf{k}+\mathbf{G})\mathbf{r}}$, $\mathbf{G}$ being 3D reciprocal-lattice vectors. The Fourier components $V_{\Delta\mathbf{K}} = <e^{i(\mathbf{k}+\mathbf{G})\mathbf{r}}|V(\mathbf{r})|e^{i(\mathbf{k}+\mathbf{G}')\mathbf{r}}>$ of the local pseudopotential $V(\mathbf{r})$ were adjustable parameters.

We start from the ideal FE case, where $V(\mathbf{r})$ is constant and equal to $V_0$ (so-called empty lattice). The corresponding calculations are plotted in Fig. 3 (a) as the $E(\mathrm{Re}k_z)$ bands (the corresponding $E(\mathrm{Im}k_z)$ bands are not shown here for brevity). Due to the absence of hybridization between the plane waves in the empty-lattice case, each $\phi_\mathbf{k}(\mathbf{r})$ contains one single plane wave corresponding to a certain $\mathbf{G}$ vector. Typical of high energies, we observe a dense multitude of bands brought in by an immense number of all $\mathbf{G}$ vectors falling into our energy region. Starting from the ultimate $V_0 = 0$ case, when the vacuum half-space is identical to the crystal one, it is obvious that only one band will couple to the photoelectron plane wave in vacuum $e^{i\mathbf{K}\mathbf{r}}$ and thus be effective in the ARPES final state, specifically, only the primary band whose plane wave – in the context of LEED often called conducting plane wave – has $\mathbf{k}+\mathbf{G}$ equal to the photoelectron $\mathbf{K}$. The whole multitude of the secondary bands, whose plane wave's $\mathbf{k}+\mathbf{G}$ is different from $\mathbf{K}$, will give no contribution to the photocurrent. In our more general case $V(\mathbf{r}) = V_0$, the $k_z$ component of the photoelectron distorts upon its escape to vacuum, and the above momentum-equality condition to identify the conducting plane wave should be cast in terms of the in-plane components as $\mathbf{k}_{//} + \mathbf{G}_{//} = \mathbf{K}_{//}$. In a formal language, these intuitive considerations can be expressed through the partial contributions of each $\phi_\mathbf{k}(\mathbf{r})$ into the total current absorbed in the sample in the LEED process, which are the so-called partial absorbed currents $T_\mathbf{k} \propto V_i \cdot \int_0^\infty |A_k \phi_k(z)|^2 dz$, with the integration extending from the crystal surface into its depth[31,32,37]. Importantly in the ARPES context, the $T_\mathbf{k}$ values multiplied by the photoemission matrix elements define the partial photocurrents emanating from the individual $\phi_\mathbf{k}(\mathbf{r})$ in the MBFS[31]. In Fig. 3(a) the calculated $T_\mathbf{k}$ are marked in blue colorscale. As expected for the empty-lattice case, $T_\mathbf{k}$ is equal to 1 for the primary (in the LEED context often called conducting) band and strictly zero for all other ones, realising the ideal FE final state containing one single plane wave. In Mahan's language, only the primary-cone photoemission is active in our ideal FE case.

We will now introduce spatial modulations of $V(\mathbf{r})$ as expressed by $V_{\Delta\mathbf{K}}$ for non-zero $\Delta\mathbf{K}$. The plane waves start to hybridise through the $V_{\Delta\mathbf{K}}$ matrix elements, and each $\phi_\mathbf{k}(\mathbf{r})$ becomes a superposition of a few plane waves as $\phi_\mathbf{k}(\mathbf{r}) = \Sigma_\mathbf{G} C_\mathbf{G} e^{i(\mathbf{k}+\mathbf{G})\mathbf{r}}$. In this case not only one but several $\phi_\mathbf{k}(\mathbf{r})$ can acquire a certain admixture of the $\mathbf{k}_{//} + \mathbf{G}_{//} = \mathbf{K}_{//}$ conducting plane wave – in the formal language, their $T_\mathbf{k}$ becomes non-zero – and give a certain contribution to the total photocurrent. Our model calculations for this case are sketched in Fig. 3 (b). The ARPES final state appears multiband in a sense that it consists of several $\phi_\mathbf{k}(\mathbf{r})$ with different $k_z$s (typically alongside the primary band) which give comparable contributions to the total ARPES signal as quantified by the corresponding $T_\mathbf{k}$. In Mahan's language, the qualitative distinction between the primary- and secondary-cone photoemission dissolves. Correspondingly, the ARPES spectra will show up several peaks corresponding to different $k_z$ or, if the separation of these $k_z$s is smaller than the intrinsic $\Delta k_z$, excessive broadening of the spectral peaks. This is exactly what we have just seen in our ARPES data on Ag(100). We note in passing that on the qualitative level the bands contributing to the photocurrent can be easily identified based on the Fourier expansion of their $\phi_k(\mathbf{r})$ which should have a substantial weight of the $\mathbf{k}_{//} + \mathbf{G}_{//} = \mathbf{K}_{//}$ conducting plane wave[50].

Whereas for the sake of physical insight we have intentionally simplified the above picture, the exact treatment of the MBFSs based on the matching approach of LEED has been developed in a series of previous works albeit limited to relatively low final-state energies[31,34,37,38]. Finally, we note that the MBFS phenomenon can also be understood within the simplified three-step model of photoemission, where the whole quantum-mechanical photoemission process is splitted into the photoexcitation of a photoelectron, its transport out of the crystal, and escape to vacuum. In this framework, the MBFSs can be viewed as resulting from multiple scattering of photoelectrons on their way out of the crystal that creates multiple Bloch-wave modes of the scattered wavefield.

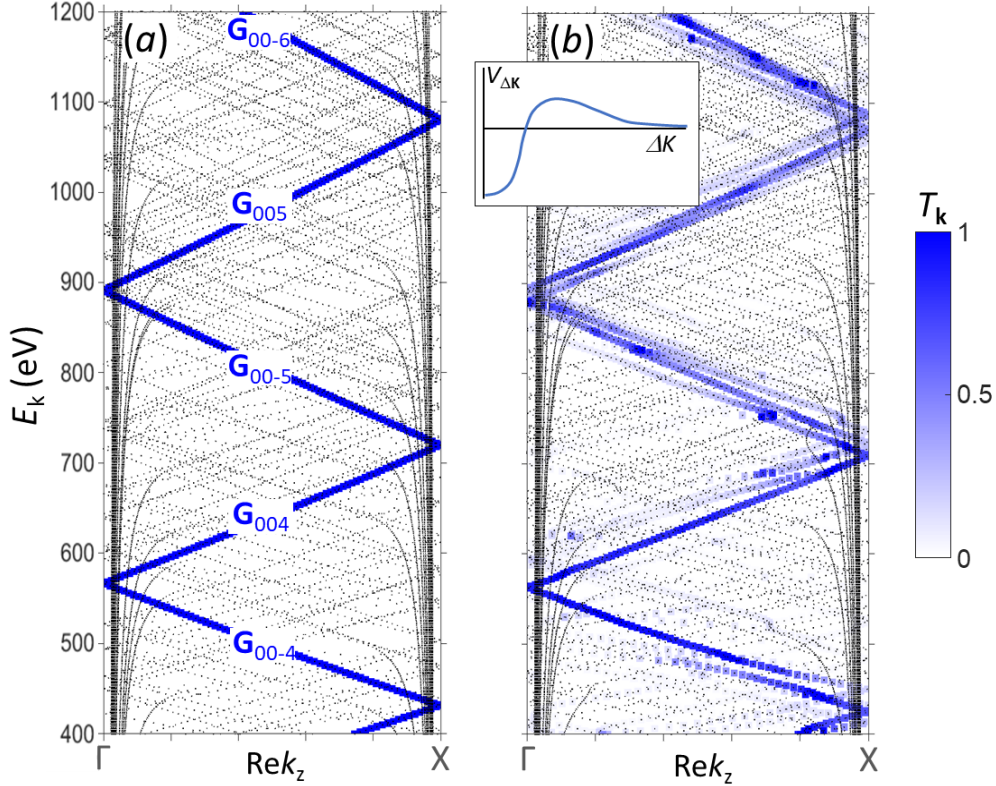

Fig. 3. Band structure of the final-state Bloch waves $E(\text{Re}k_z)$ in a model fcc crystal along the ΓX direction (a) in the empty-lattice case $V(\mathbf{r}) = V_0$ and (b) with a more realistic spatially modulated pseudopotential, sketched in the insert. The dense multitude of bands is formed by an immense number of **G** vectors falling into our high-energy region. The contributions of each band into the total photocurrent are quantified by $T_k$ (blue colorscale). Whereas in the first case the photocurrent emanates from one single FE band (marked with the corresponding **G** vectors), in the second case it may distribute over a few bands alongside the FE dispersion, which form a MBFS incorporating a few $k_z$s.

Whereas the effects of MBFSs have already been established at low excitation energies, their survival in high-energy ARPES might seem puzzling. In a naive way of thinking, photoelectrons with energies much higher than the modulations of $V(\mathbf{r})$ should not feel them, recovering the FE case with one single $\phi_k(\mathbf{r})$. However, $V_{\Delta\mathbf{K}}$ as the strength of hybridization between two plane waves depends, somewhat counter-intuitively, not on energy but rather on $\Delta\mathbf{K}$ between them. As sketched in the insert in Fig. 3 (b), $V_{\Delta\mathbf{K}}$ typically has its maximal negative value at $\Delta K = 0$ (which is the $V_0$), and with increase of $\Delta K$ sharply rises and then asymptotically vanishes. Importantly, however high the energy is, the multitude of the plane waves always contains pairs of those whose $\Delta K$ is small. The corresponding bands can be identified by close dispersions. For such pairs $V_{\Delta\mathbf{K}}$ is large, giving rise to their strong hybridization. Importantly, all bands hybridising with the $\mathbf{k}_{//} + \mathbf{G}_{//} = \mathbf{K}_{//}$ plane wave will receive non-zero $T_k$ and thus contribute to the total photocurrent, as shown in Fig. 3 (b). This forms the MBFSs that should survive even at high energies.

## Effect of MBFSs on the spectral structure

We will now follow in more detail how the MBFSs affect the ARPES spectra. As an example, we will analyse the experimental $k_z$-MDC from Fig. 2(d) in the region of the X point at $hv \sim 1100$ eV, reproduced in Fig. 4 (with the linear background subtracted). Within the FE approximation, we might expect to observe here two Lorentzian peaks, placed symmetrically around the X point and broadened by the same intrinsic $\Delta k_z$. However, the $k_z$-MDC shows three distinct peaks *A-C*, with the peak *B* coming from a final-state band falling beyond the FE approximation. Moreover, Lorentzian fitting of the peaks finds that whereas the peak *C* has a relatively small width of 0.11 Å$^{-1}$, the widths of the peaks *A* and *B* are more than twice larger, 0.30 and 0.32 Å$^{-1}$, respectively. The picture of MBFSs neatly explains this observation, suggesting that whereas the peak *C* is formed by a final state having one dominant $k_z$ contribution, and the peaks *A* and *B* by final states incorporating a multitude of $k_z$s separated less than $\Delta k_z$. Whereas it is generally believed that the intrinsic broadening of the ARPES peaks in $k_z$ is determined exclusively by finite $\lambda_{PE}$ the photoelectron mean free path, our example demonstrates that the multiband final-state composition may not only create additional spectral peaks but also be an important factor of their broadening additional to $\lambda_{PE}$.

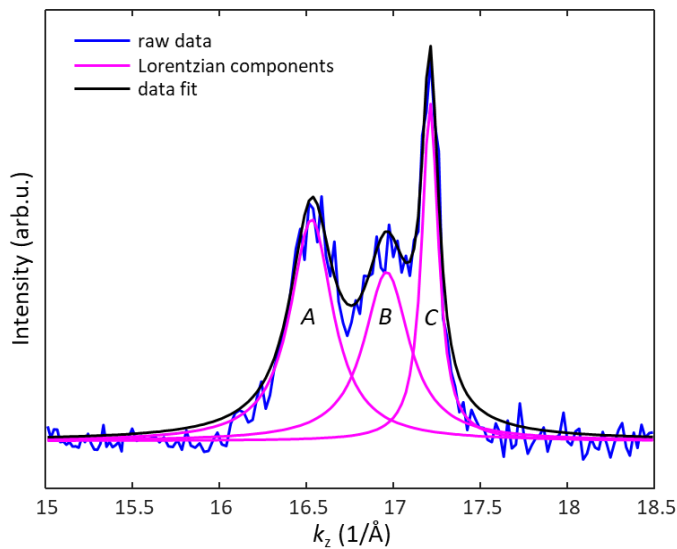

Fig. 4. $k_z$-MDC at $E_F$ from Fig. 2(d) in the $hv$ region around 1100 eV (vicinity of the X point) decomposed in three Lorentzians. The presence of the peak *B* and the larger broadening of the peaks *A* and *B* compared to *C* are caused by MBFSs.

Intriguingly, however, we note that even the narrowest peak *C* is almost twice broader than $\Delta k_z \sim 0.065$ Å$^{-1}$ expected from $\lambda_{PE} \sim 15.5$ Å suggested by the TPP-2M formula [51] well-established in XPS and Auger electron spectroscopy. One explanation might be that already the peak *C* would incorporate multiple final-state bands with smaller $k_z$ separation compared to other two peaks. Another explanation would trace back to quasielastic electron-electron or electron-phonon scattering, which would increase with energy owing to the increase of the phase-space volume available for such scattering. Altering **k** of photoelectrons, it should destroy the coherence of photoelectrons and thus reduce $\lambda_{PE}$ as reflected in the observed $\Delta k_z$. At the same time, the quasielastic scattering should have only a little effect on attenuation of the **k**-integrated signal of the core-level or intrinsically incoherent Auger electrons. In other words, the effective $\lambda_{PE}$ in ARPES should be smaller than that in XPS/Auger spectroscopy, described by the TPP-2M and related formalism. Such intriguing fundamental physics certainly deserves further investigation.

## MBFS phenomena through various materials

The phenomenon of MBFSs surviving at high excitation energies is certainly not restricted to Ag only and, strengthening with the strength of $V(\mathbf{r})$ modulations, should be fairly general over various materials. Even for Al the paradigm FE metal, astonishingly, such MBFSs can be detected at least up to excitation energies of a few hundreds of eV[14,42]. Quite commonly the MBFS effects at high energies are observed in van-der-Waals materials such as $MoTe_2$[52], which should be connected with a large modulation of $V(\mathbf{r})$ across the van-der-Waals gap.

Another vivid example of the MBFS effects is the soft-X-ray ARPES data for GaN presented in Fig. 5, compiled from the previously published results on AlN/GaN(1000) heterostructures[13]. The panel (a) shows the ARPES spectral structure plot expected from the DFT valence bands and FE final states with $V_0 = 5$ eV. With the non-symmorphic space group of bulk GaN, the ARPES dispersions allowed by the dipole selection rules (though in our case somewhat relaxed due to the band bending in GaN) are marked bold. The panels (b,c) present the experimental out-of-plane ARPES dispersions measured at $k_x$ in two formally equivalent $\overline{\Gamma}$ points of the surface BZ, $\overline{\Gamma}_0$ in the first and $\overline{\Gamma}_1$ in the second zone. As expected because of weaker electron screening of the atomic potential and thus sharper modulations of $V(\mathbf{r})$ in the covalent GaN compared to the metallic Ag, the deviations of experimental dispersions from the predictions of the FE approximation are much stronger than for Ag. One can clearly see the MBFSs where the individual bands (marked by arrows at their top) are separated in $k_z$ more than the intrinsic $\Delta k_z$ broadening. In the multitude of the experimental ARPES dispersions, one can identify the one which can be associated with the primary-cone photoemission (bold arrows) although in the $\overline{\Gamma}_0$ data this band cannot be traced below 1000 eV. Remarkably, for the same initial-state $E(\mathbf{k})$ the ARPES dispersions measured at the $\overline{\Gamma}_0$ and $\overline{\Gamma}_1$ points appear completely different, identifying different final-state bands selected from the continuum of all unoccupied states available for given final-state energy and $\mathbf{K}_{//}$. These bands are identified by their leading plane-wave component to have $\mathbf{k}_{//} + \mathbf{G}_{//} = \mathbf{K}_{//}$, where $\mathbf{K}_{//}$ of the photoelectron changes between the surface BZs[32].

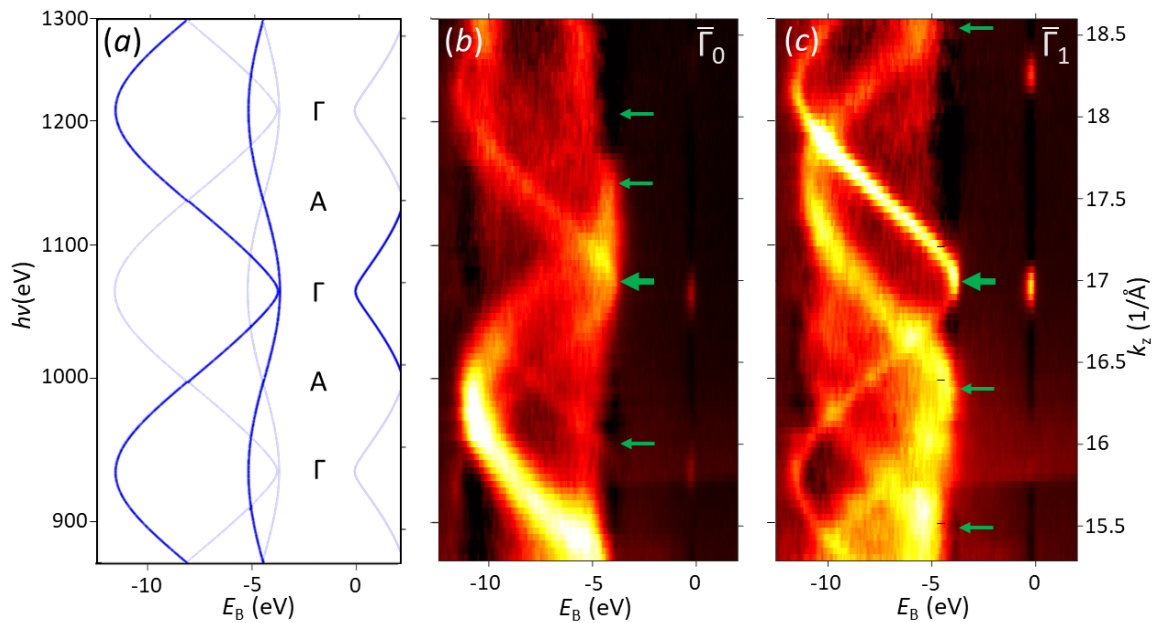

Fig. 5. Out-of-plane ARPES dispersions for GaN(1000): (a) Expected from the DFT valence bands and FE final states with $V_0 = 5$ eV. With the non-symmorphic space group of bulk GaN, the dispersions allowed by the dipole selection rules are shown bold; (b,c) Measured at $k_x = 0$ projecting onto the $\overline{\Gamma}_0$ and $\overline{\Gamma}_1$ points over two BZs. The experiment clearly resolves individual final-state bands (marked by arrows) whose separation in $k_z$ is larger than the intrinsic $\Delta k_z$ broadening.

The high-energy final states in Si are a counter-example though. Fig. 6 presents soft-X-ray ARPES data on a few-nm thick layer of Si(100) n-doped with As[53] as the out-of-plane band dispersions (b) and iso-$E_B$ contours (c), respectively. The panel (a) shows the ARPES spectral structure plot expected from the DFT-GGA calculated valence bands and FE final states with $V_0$ = 10 eV, with the bold lines indicating the dispersions allowed by the selection rules (for in-depth discussion see Ref. 54). Because of the covalent character of Si, one might again expect that the non-FE effects here would be comparable to those for GaN and in any case stronger than for the metallic Ag. Contrary to such expectations, however, the experimental data in (b,c) does not show any clear signatures of the MBFSs in Si in the shown ($E_k$,**k**) region, although at low excitation energies they are profound[35]. At the moment we can not decipher any simple arguments that would relate the strength of the non-FE effects in the high-energy electron states to any obvious electronic-structure parameters of various materials.

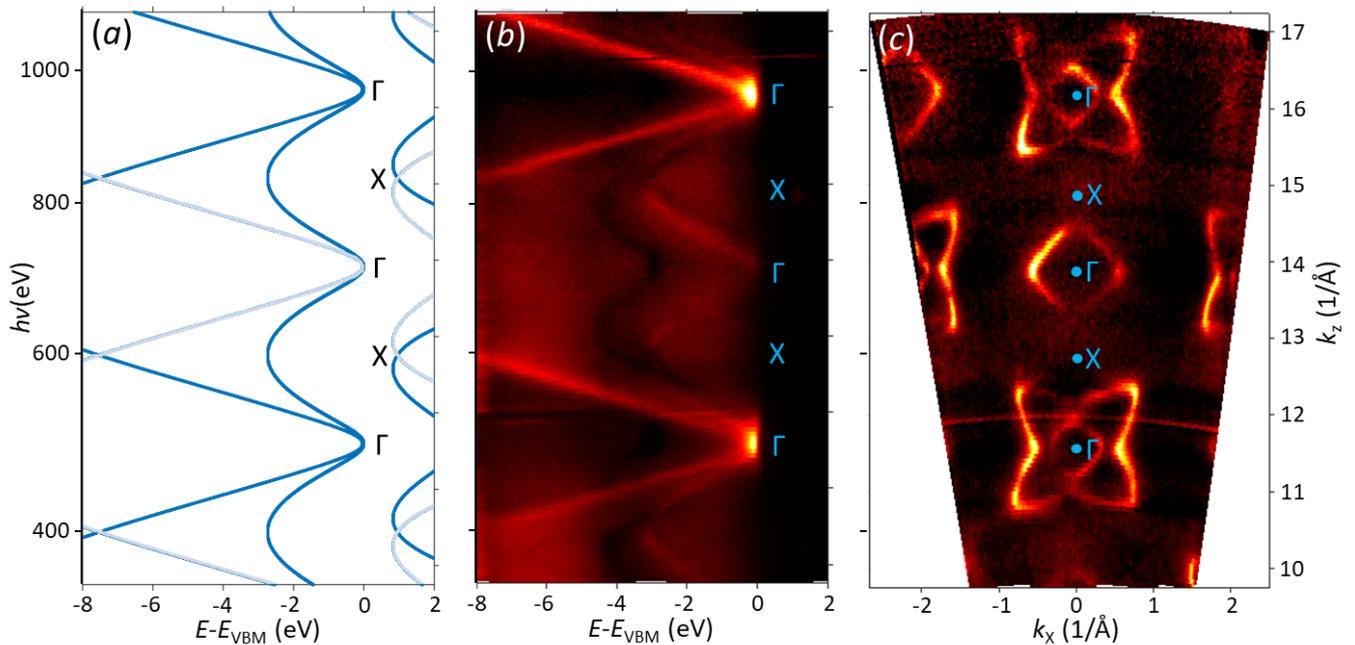

Fig. 6. Out-of-plane ARPES data for Si(100): (a) ARPES dispersions expected from the DFT valence bands and FE final states with $V_0$ = 10 eV, with dispersions allowed by the selection rules shown bold; (b) Experimental band dispersions and (c) iso-$E_B$ contours in 2 eV below the valence-band maximum. No clear signatures of the MBFSs can be identified in these data.

## Non-FE effects beyond ARPES

The non-FE effects in high-energy electron states such as MBFS manifest themselves not only in the ARPES dispersions. Another manifestation will be the circular dichroism in the angular distribution of photoelectrons (CDAD) that necessitates that the final-state wavefunctions deviate from the free-electron plane waves[55,56]. The CDAD has indeed been observed already in the early soft-X-ray ARPES study on Ag(100)[39]. Another example is the orbital tomography of adsorbed molecules (see, for example, Refs. 57–59) which takes advantage of the Fourier relation between the angle distribution of photoelectrons and electron density of the valence electron orbitals. The non-FE effects introduce additional plane-wave components in the final states, calling for refinement of the straightforward Fourier-transform processing of the experimental data[59]. Beyond ARPES, the very fact of electron diffraction at crystalline surfaces identifies non-FE effects in the electron states in the crystal, because otherwise the incident electrons would upon entering the crystal follow the same FE wavefunction and thus would not reflect. The Reflection High-Energy Electron Diffraction (RHEED) evidences that the non-FE effects survive even in the energy range of a few tens of keV, when Δ**K** between the incident and diffracted plane waves is small and thus the corresponding $V_{\Delta \mathbf{K}}$ large. These considerations suggest that the MBFSs should survive even in hard-X-ray ARPES, waiting for a direct experimental observation.

Finally, we should point out that the coherent photoemission process underlying the ARPES experiment discussed above (as well as the orbital tomography) is fundamentally different to the essentially incoherent process of X-ray photoelectron diffraction (XPD) (see, for example, the reviews[60–62]. In the first case, all photoelectron emitters (atoms) throughout the crystal surface region within the depth $\lambda_{PE}$ are coherent – or entangled, in the modern quantum mechanics discourse – and emit a coherent photoelectron wavefield characterised by a well-defined **k**. The resulting ARPES intensity as a function of $E_k$ and $\theta$ bears sharp structures reflecting, through the momentum conservation, the **k**-resolved band structure of the valence states. In the XPD, other way around, the coherence between the emitters throughout the surface region is lost. This takes place, for example, for isolated impurity atoms or adsorbed molecules, localised core levels, where the initial-state wavefunctions at different atoms are decoupled from each other, or when the coherence of photoelectrons is broken by thermal or defect scattering, or when the signal from certain valence-band states, like $d$-states, is integrated in energy[63,64]. The result is that each photoelectron emitter creates scattered waves within a sphere of the radius $\lambda_{PE}$, which interfere with each other incoherently with the waves emanating from another emitter. Typical of diffraction with a few interfering rays, the resulting XPD intensity distribution as a function of $E_k$ and $\theta$ is fairly smooth, and reflects the local atomic structure. With $E_k$ increasing into the hard-X-ray energy range, $\lambda_{PE}$ and thereby the number of coherently scattered waves increases. This forms sharp Kikuchi-like structures in the XPD angular distribution, reflecting the long-range atomic structure[62]. In any case, the XPD stays incoherent between the emitters. This fundamental difference between the coherent photoemission and incoherent XPD processes is stressed, for example, by the fact that in the first case the photoelectron angular distribution follows $\mathbf{p}_{ph}$, shifting with $h\nu$, and in the second case it is insensitive to $\mathbf{p}_{ph}$[19].

# Conclusion

Our analysis of extensive soft-X-ray ARPES data on the Ag metal has demonstrated that even at high excitation energies the photoemission final states may, intriguingly, in some energy and **k**-space regions feature pronounced multiband composition beyond the conventional FE approximation. The corresponding Bloch waves have different $k_z$ momenta, typically alongside the FE dispersion, and give comparable contribution to the ARPES spectra. Using empirical-pseudopotential simulation of the final states, where these contributions were quantified as proportional to the partial current in each Bloch wave determined within the wavefunction-matching formalism of LEED, we have demonstrated that the MBFSs appear due to hybridization of plane waves through low-**K** components of the crystal potential. Depending on the $k_z$ separation of the individual Bloch waves, the MBFSs give rise to multiple ARPES peaks from 3D valence-band dispersions or become an important factor of their broadening in addition to the intrinsic $\Delta k_z$ broadening due to the finite $\lambda_{PE}$. From the first principles, these effects can be described by one-step ARPES calculations with the final states treated within the multiple-scattering or Bloch-wave approaches. Although our KKR-based calculations on Ag were able to qualitatively describe the experimental results, further theoretical effort is required to achieve a quantitative agreement at high excitation energies. Besides Ag, the MBFS phenomena are observed, for example, in previous soft-X-ray data on the covalent GaN and even Al, the paradigm FE metal. They are surprisingly weak, however, for the covalent Si. The MBFS phenomenon, typically strengthening with the sharpness of the crystal-potential modulations, should be fairly general over a wide range of materials and excitation energies even into the hard-X-ray range.

# Methods

## Experiment

The experiments were performed at the soft-X-ray ARPES facility[65] installed at the high-resolution ADRESS beamline[66] of the Swiss Light Source, Paul Scherrer Institute, Switzerland. X-rays irradiated the sample with a flux of ~$10^{13}$ photons/s at a grazing-incidence angle of 20°. A single crystal of Ag(100) (MaTecK) was cleaned by a few cycles of Ar ion sputtering/annealing. The sample was cooled down to ~12K in order to quench relaxation of **k**-conservation due to thermal motion of the atoms[67], with the coherent spectral fraction enhanced by subtracting the angle-integrated spectrum scaled under the condition of non-negativity of the remaining spectral weight. The measurements were performed with p-polarised X-rays at a combined energy resolution varying from ~50 to 180 meV when going from $hv$ = 300 to 1300 eV, which is about twice better than in the first soft-X-ray ARPES study on Ag(100)[39]. The FS maps were integrated over an $E_B$ window from -75 to 25 meV relative to $E_F$. Angular resolution of the analyzer PHOIBOS-150 was ~0.1°. Other relevant experimental details, including the conversion of $E_k$ and emission angle $\theta$ to **k**, corrected for **p**$_{ph}$, can be found elsewhere[65]. The data on GaN and Si from the previous ARPES works, discussed below, were taken under the same experimental conditions, but with the energy resolution relaxed to ~80 to 250 meV in the same $hv$ range.

## Calculations

In our simulations of the photoemission final states, the use of an empirical local pseudopotential has allowed reduction of the secular equation on complex $k_z$ to an eigenvalue problem for a complex non-Hermitian matrix[17,45]. For the energy range of our simulation extending to 1200 eV, the basis set included all plane waves below an energy cutoff of 1800 eV. The inner potential $V_0$ was set to 10 eV, all $V_{\Delta K}$ to 5 eV for $\Delta K^2$ < 48 and to zero for larger $\Delta K^2$, and $V_i$ to 5 eV. The accuracy of the calculations was controlled via the current conservation generalised for non-zero $V_i$ on the crystal side. For our qualitative analysis of the final states, no attempt has been made to fit these parameters to our particular case. Details of the calculations can be found elsewhere[32].

The first-principles ARPES calculations were performed using the SPR-KKR package[68] relying on the multiple scattering theory using the Korringa-Kohn-Rostoker (KKR) method. The ground-state properties of the Ag(001) surface were derived from density-functional-theory (DFT) calculations within the local-density approximation (LDA) carried out with full potential. The ARPES spectra were calculated within the one-step model of photoemission in the spin-density-matrix formulation[69] taking into account all aspects of the photoemission process for the actual experiment including **p**$_{ph}$, matrix elements and final states constructed as the time-reversed LEED states. Taking into advantage the predominance of forward scattering at $E_k$ above ~400 eV[70] the calculations used the single-site scattering approximation. The final-state damping was described via constant $V_i$ = 3 eV set to reproduce $\lambda_{PE}$ = 10.2 Å at $E_k$ = 600 eV given by the TPP-2M formula[51]. For further computational details see Supplemental Material. The main paper presents the results obtained with FE final states, and the effects of multiple-scattering final states and various computational approximations are discussed in Supplemental Material.

# Data availability

The raw and derived data presented are available from the corresponding authors upon a reasonable request.

# References

1. Weng, H., Fang, C., Fang, Z., Andrei Bernevig, B. & Dai, X. Weyl Semimetal Phase in Noncentrosymmetric Transition-Metal Monophosphides. *Physical Review X* **5**, 011029 (2015).

2. Lv, B. Q. *et al.* Observation of Weyl nodes in TaAs. *Nat. Phys.* **11**, 724 (2015).

3. Schröter, N. B. M. *et al.* Chiral topological semimetal with multifold band crossings and long Fermi arcs. *Nat. Phys.* **15**, 759 (2019).

4. Schröter, N. B. M. *et al.* Observation and control of maximal Chern numbers in a chiral topological semimetal. *Science* **369**, 179 (2020).

5. Armitage, N. P., Mele, E. J. & Vishwanath, A. Weyl and Dirac semimetals in three-dimensional solids. *Rev. Mod. Phys.* **90**, 015001 (2018).

6. Yan, B. & Felser, C. Topological Materials: Weyl Semimetals. *Annu. Rev. Condens. Matter Phys.* **8**, 337 (2017).

7. Lv, B., Qian, T. & Ding, H. Angle-resolved photoemission spectroscopy and its application to topological materials. *Nature Reviews Physics* **1**, 609 (2019).

8. Hasan, M. Z. *et al.* Weyl, Dirac and high-fold chiral fermions in topological quantum matter. *Nature Reviews Materials* **6**, 784 (2021).

9. Strocov, V. N. *et al.* Three-dimensional electron realm in $VSe_2$ by soft-x-ray photoelectron spectroscopy: Origin of charge-density waves. *Phys. Rev. Lett.* **109**, 086401 (2012).

10. Weber, F. *et al.* Three-dimensional Fermi surface of $2H-NbSe_2$ : Implications for the mechanism of charge density waves. *Physical Review B* **97** 235122 (2018).

11. Wang, Z. *et al.* Three-dimensional charge density wave observed by angle-resolved photoemission spectroscopy in $1T-VSe_2$. *Phys. Rev. B Condens. Matter* **104**, 155134 (2021).

12. King, P. D. C. *et al.* Surface band-gap narrowing in quantized electron accumulation layers. *Phys. Rev. Lett.* **104**, 256803 (2010).

13. Lev, L. L. *et al.* k-space imaging of anisotropic 2D electron gas in GaN/GaAlN high-electron-mobility transistor heterostructures. *Nature Communications* **9**, 2653 (2018).

14. Strocov, V. N. Photoemission response of 2D electron states. *Journal of Electron Spectroscopy and Related Phenomena* vol. **229**, 100 (2018).

15. Moser, S. *et al.* How to extract the surface potential profile from the ARPES signature of a 2DEG. *J. Electron Spectrosc. Relat. Phenom.* **225**, 16 (2018).

16. Schuwalow, S. *et al.* Band structure extraction at hybrid narrow-gap semiconductor-metal interfaces. *Adv. Sci.* **8**, 2003087 (2021).

17. Smith, D. L. & Mailhiot, C. Theory of semiconductor superlattice electronic structure. *Reviews of Modern Physics* vol. **62**, 173 (1990).

18. Husanu, M.-A. *et al.* Electron-polaron dichotomy of charge carriers in perovskite oxides. *Communications Physics* **3**, 62 (2020).

19. Berner, G. *et al.* Dimensionality-tuned electronic structure of nickelate superlattices explored by soft-x-ray angle-resolved photoelectron spectroscopy. *Physical Review B* **92**, 125130 (2015).

20. Schütz, P. *et al.* Dimensionality-driven metal-insulator transition in spin-orbit-coupled $SrIrO_3$. *Phys. Rev. Lett.* **119**, 256404 (2017).

21. Tang, F. *et al.* Three-dimensional quantum Hall effect and metal–insulator transition in $ZrTe_5$. *Nature* **569**, 537 (2019).

22. Suga, S. & Tusche, C. Photoelectron spectroscopy in a wide hv region from 6eV to 8keV with full momentum and spin resolution. *Journal of Electron Spectroscopy and Related Phenomena* **200**, 119 (2015).

23. Fadley, C. S. Looking Deeper: Angle-Resolved Photoemission with Soft and Hard X-rays. *Synchrotron Radiation News* **25**, 26 (2012).

24. Gray, A. X. *et al.* Bulk electronic structure of the dilute magnetic semiconductor $Ga_{1-x}Mn_xAs$ through hard X-ray angle-resolved photoemission. *Nature Materials* **11**, 957 (2012).

25. Strocov, V. N. *et al.* Soft-X-ray ARPES at the Swiss Light Source: From 3D Materials to Buried Interfaces and Impurities. *Synchrotron Radiation News* **27**, 31 (2014).

26. Strocov, V. N. *et al.* **k**-resolved electronic structure of buried heterostructure and impurity systems by soft-X-ray ARPES. *Journal of Electron Spectroscopy and Related Phenomena* **236**, 1 (2019).

27. Powell, C. J. & Jablonski, A. Surface Sensitivity of Auger-Electron Spectroscopy and X-ray Photoelectron Spectroscopy. *Journal of Surface Analysis* **17**, 170 (2011).

28. Strocov, V. N. Intrinsic accuracy in 3-dimensional photoemission band mapping. *Journal of Electron Spectroscopy and Related Phenomena* **130**, 65 (2003).

29. Lindgren, S. Å., Walldén, L., Rundgren, J., Westrin, P. & Neve, J. Structure of Cu(111)p(2×2)Cs determined by low-energy electron diffraction. *Physical Review B* **28**, 6707 (1983).

30. Rundgren, J. Optimized surface-slab excited-state muffin-tin potential and surface core level shifts. *Physical Review B* **68**, 125405 (2003).

31. Strocov, V. N., Starnberg, H. I. & Nilsson, P. O. Excited-state bands of Cu determined by VLEED band fitting and their implications for photoemission. *Physical Review B* **56**, 1717 (1997).

32. Strocov, V. N. *et al.* Three-dimensional band mapping by angle-dependent very-low-energy electron diffraction and photoemission: Methodology and application to Cu. *Physical Review B* **63**, 205108 (2001).

33. Krasovskii, E. E. *et al.* Photoemission from Al(100) and (111): Experiment and *ab initio* theory. *Physical Review B* **78**, 165406 (2008).

34. Krasovskii, E. E. Character of the outgoing wave in soft x-ray photoemission. *Physical Review B* **102**, 245139 (2020).

35. Strocov, V. N. *et al.* Very-low-energy electron diffraction on the H-terminated Si(111) surface: Ab initio pseudopotential analysis. *Phys. Rev. B Condens. Matter* **59**, R5296 (1999).

36. Strocov, V. N., Starnberg, H. I., Nilsson, P. O., Brauer, H. E. & Holleboom, L. J. New Method for Absolute Band Structure Determination by Combining Photoemission with Very-Low-Energy Electron Diffraction: Application to Layered $VSe_2$. *Physical Review Letters* **79**, 467 (1997).

37. Strocov, V. N. *et al.* Three-dimensional band structure of layered $TiTe_2$: Photoemission final-state effects. *Physical Review B* **74**, 195125 (2006).

38. Krasovskii, E. E. *et al.* Band mapping in the one-step photoemission theory: Multi-Bloch-wave structure of final states and interference effects. *Physical Review B* **75**, 045432 (2007).

39. Venturini, F., Minár, J., Braun, J., Ebert, H. & Brookes, N. B. Soft x-ray angle-resolved photoemission spectroscopy on Ag(001): Band mapping, photon momentum effects, and circular dichroism. *Physical Review B* **77**, 045126 (2008).

40. Strocov, V. N., Claessen, R., Aryasetiawan, F., Blaha, P. & Nilsson, P. O. Band- and **k**-dependent self-energy effects in the unoccupied and occupied quasiparticle band structure of Cu. *Physical Review B* **66**, 195104 (2002).

41. Mahan, G. D. Theory of Photoemission in Simple Metals. *Physical Review B* **2**, 4334 (1970).

42. Hofmann, P. *et al.* Unexpected surface sensitivity at high energies in angle-resolved photoemission. *Physical Review B* **66**, 245422 (2002).

43. Feibelman, P. J. & Eastman, D. E. Photoemission spectroscopy—Correspondence between quantum theory and experimental phenomenology. *Physical Review B* **10**, 4932 (1974).

44. Capart, G. Band structure calculations of low energy electron diffraction at crystal surfaces. *Surface Science* **13**, 361 (1969).

45. Pendry, J. B. The application of pseudopotentials to low-energy electron diffraction III: The simplifying effect of inelastic scattering. *Journal of Physics C: Solid State Physics* **2**, 2283 (1969).

46. Dederichs, P. H. Dynamical Diffraction Theory by Optical Potential Methods. *Solid State Physics* **27** (1972) eds. H. Ehrenreich, F. Seitz & D. Turnbull (New York: Academic) p. 136.

47. Barrett, N., Krasovskii, E. E., Themlin, J.-M. & Strocov, V. N. Elastic scattering effects in the electron mean free path in a graphite overlayer studied by photoelectron spectroscopy and LEED. *Physical Review B* **71**, 035427 (2005).

48. Krasovskii, E. E. & Schattke, W. Angle-Resolved Photoemission from Surface States. *Physical Review Letters* **93**, 027601 (2004).

49. Heine, V. On the General Theory of Surface States and Scattering of Electrons in Solids. *Proceedings of the Physical Society* **81**, 300 (1963).

50. Strocov, V. N. On qualitative analysis of the upper band effects in very-low-energy electron diffraction and photoemission. *Solid State Communications* **106**, 101 (1998).

51. Tanuma, S., Powell, C. J. & Penn, D. R. Proposed formula for electron inelastic mean free paths based on calculations for 31 materials. *Surface Science* **192**, L849 (1987).

52. J. Krieger *et al*. Unpublished (2020)

53. Stock, T. J. Z. *et al.* Atomic-Scale Patterning of Arsenic in Silicon by Scanning Tunneling Microscopy. *ACS Nano* **14**, 3316 (2020).

54. P. Constantinou *et al*. Unpublished (2020)

55. Moser, S. An experimentalist's guide to the matrix element in angle resolved photoemission. *Journal of Electron Spectroscopy and Related Phenomena* **214**, 29 (2017).

56. Fedchenko, O. *et al.* 4D texture of circular dichroism in soft-x-ray photoemission from tungsten. *New Journal of Physics* **21**, 013017 (2019).

57. Puschnig, P. *et al.* Reconstruction of Molecular Orbital Densities from Photoemission Data. *Science*

**326**, 702 (2009).

58. Kliuiev, P. *et al.* Combined orbital tomography study of multi-configurational molecular adsorbate systems. *Nature Communications* **10**, 5255 (2019).

59. Bradshaw, A. M. & Woodruff, D. P. Molecular orbital tomography for adsorbed molecules: is a correct description of the final state really unimportant? *New Journal of Physics* **17**, 013033 (2015).

60. Fadley, C. S., Van Hove, M. A., Hussain, Z. & Kaduwela, A. P. Photoelectron diffraction: new dimensions in space, time, and spin. *Journal of Electron Spectroscopy and Related Phenomena* **75**, 273 (1995).

61. Woodruff, D. Adsorbate structure determination using photoelectron diffraction: Methods and applications. *Surface Science Reports* **62**, 1 (2007).

62. Fedchenko, O., Winkelmann, A. & Schönhense, G. Structure Analysis Using Time-of-Flight Momentum Microscopy with Hard X-rays: Status and Prospects. *Journal of the Physical Society of Japan* **91**, 091006 (2022).

63. Osterwalder, J., Greber, T., Hüfner, S. & Schlapbach, L. X-ray photoelectron diffraction from a free-electron-metal valence band: Evidence for hole-state localization. *Physical Review Letters* **64**, 2683 (1990).

64. Osterwalder, J., Greber, T., Aebi, P., Fasel, R. & Schlapbach, L. Final-state scattering in angle-resolved ultraviolet photoemission from copper. *Physical Review B* **53**, 10209 (1996).

65. Strocov, V. N. *et al.* Soft-X-ray ARPES facility at the ADRESS beamline of the SLS: concepts, technical realisation and scientific applications. *Journal of Synchrotron Radiation* **21**, 32 (2014).

66. Strocov, V. N. *et al.* High-resolution soft X-ray beamline ADRESS at the Swiss Light Source for resonant inelastic X-ray scattering and angle-resolved photoelectron spectroscopies. *Journal of Synchrotron Radiation* **17**, 631 (2010).

67. Braun, J. *et al.* Exploring the XPS limit in soft and hard x-ray angle-resolved photoemission using a temperature-dependent one-step theory. *Physical Review B* **88**, 205409 (2013).

68. Ebert, H., Ködderitzsch, D. & Minár, J. Calculating condensed matter properties using the KKR-Green's function method—recent developments and applications. *Reports on Progress in Physics* **74**, 096501 (2011).

69. Braun, J., Minár, J. & Ebert, H. Correlation, temperature and disorder: Recent developments in the one-step description of angle-resolved photoemission. *Physics Reports* **740**, 1 (2018).

70. Sébilleau, D., Tricot S. & Koide, A. Unpublished (2022)

# Acknowledgements

V.N.S. thanks E.E. Krasovskii for illuminating discussions and critical reading of the manuscript, and J. H. Dil for valuable exchange on physics of XPD. J.M. is grateful to D. Sébilleau, S. Tricot and A. Koide for sharing their scattering-amplitude calculations. The authors thank N.J. Curson and S.R. Schofield for giving access to Si samples prepared at University College London. J.M. and L.N. acknowledge the support of the Czech Ministry of Education, Youth and Sports via the grant CEDAMNF CZ.02.1.01/0.0/0.0/15_003/0000358 and the support from GACR Project No. 2018725S. L.L.L. acknowledges the financial support from the Ministry of Science and Higher Education of the Russian Federation, grant #075-11-2021-086. T.J.Z.S. acknowledges the financial support of the Engineering and

Physical Sciences Research Council (grants nos. EP/R034540/1, EP/W000520/1), and Innovate UK (grant no. 75574).

## Author contributions

V.N.S. and J.M. conceived the SX-ARPES experiment at the Swiss Light Source. V.N.S., L.L.L., F.A. and L.N. performed the experiment supported by T.S. T.J.Z.S. fabricated the thin-film Si samples. V.N.S. processed and interpreted the data, and performed computational simulation of the final states supported by P.C. J.M. performed the first-principles ARPES calculations. V.N.S. wrote the manuscript with contributions from J.M., L.L.L., P.C., T.J.Z.S. and J.O. All authors discussed the results, interpretations, and scientific concepts.

## Competing interests

The authors declare no competing interests.

# Supplemental Material: KKR calculations with multiple-scattering final states

## Computational scheme

In the first step of our theoretical investigations, we performed self-consistent electronic structure calculations within the ab-initio framework of the spin-density functional theory in order to generate the self-consistent-field (SCF) potential for further photoemission calculations. The LDA potential of Vosko et al. was used[1]. The electronic structure of semi-infinite crystal was calculated within the relativistic multiple scattering approach using the Green's function Korringa-Kohn-Rostoker (KKR) formalism in the tight binding mode[2]. The experimental lattice constant ($a$ = 4.09 Å) was used. In order to achieve precise description of the most subtle details of the SCF potential, important for photoemission at high excitation energies, the multipole expansion of the Green's function employed an unusually large angular-momentum cutoff $l_{max}$ of 5. In addition, a large number of **k**-points (36x36x36) in the first surface BZ was used. The self-consistent calculations have been performed in two modi, within the so-called atomic sphere approximation (ASA) and in the full potential (FP) mode.

The obtained SCF potential was used for the photoemission calculations within the one-step model. The final states (the time-reversed LEED state) were treated using the so-called layer KKR technique[3], allowing accurate description of these states in a wide $hv$ range starting from 6 eV up to several keV. To ensure the convergence of the multiple scattering between the layers, our calculations used a plane-wave basis where the number of the surface reciprocal lattice vectors **g** was increased to 147 instead of the default value 37[2]. Another important ingredient of the multiple-scattering calculations is an accurate description of the kinematic and dynamic effects in both initial and final states. For the latter, the dynamic effects are taken into account via the X-matrix[4] which represents the energy-dependent multiple scattering within a single layer. Whereas in VUV-ARPES the kinematic and dynamic effects are comparable, in the soft- and hard-X-ray regime the dynamic effects weaken, whereby the X-matrix approaches zero, leading to the so-called single-site scattering approximation. Another important parameter in the description of multiple scattering is connected with the expansion of all physical quantities in terms of angular momentum $l$, i.e. using the Bauer's identity to represent plane waves (scattering between the layers) and spherical waves (inside the layer). These expansions involve a summation over $l$ that must be truncated at a certain value $l_{max}$. In this context, the increase of $l_{max}$ should be viewed rather as an extension of the basis set for accurate description of the multiple scattering than physically meaningful $l$-channels in the scattering process. A simple assessment of $l_{max}$ can be obtained from the radial Schrödinger equation where, in order to scatter on the spherical potential, the electron must first overcome the centrifugal barrier $l(l+1)/a^2$ ($a$ is the atomic radius). This implies only the partial waves, whose $l$ satisfies the inequality $k^2 > l(l+1)/a^2$, should be included into the $l$-expansion. The higher $E_k$, the larger $l_{max}$ needs to be used (for the detailed explanation see Ref. 5). For $E_k$ in the range 300-1300eV, considered here, $l_{max}$ falls between 4 and 5. The calculations have been performed for a finite temperature of 20K leading to an additional final-state **k**-broadening, increasing with $hv$[6].

# Effect of various approximations for the multiple-scattering process

We have made an effort to elaborate our ARPES calculations towards their quantitative agreement with the experiment in a few successive steps:

– Fig. S1 shows the results obtained with multiple-scattering final states, as opposed to the FE final states used for the calculations in Figs. 1 and 2 in the main text, under successive refinements of the computational approximations:

– The results in Fig. S1(a) were obtained within the ASA and $l_{max}$ = 3. Due to the non-FE effects described by the multiple-scattering final states, they already show spectral structures due to the MBFSs (such as where marked by magenta arrows) although mostly on the low-energy end of our $h\nu$ range and not exactly in the same **k**-space regions compared to the experiment;

– The inclusion of warping of the potential in the interstitial and surface regions within the FP scheme[7], Fig. S1(b), does not result in any significant improvement in our case of Ag. Nevertheless, we anticipate that the accurate FP will be crucial for more open crystal structures, covalent materials, van-der-Waals materials, etc. where the potential modulations are sharper[7]. As we have seen in the present work, their accurate description should be particularly important at high $E_k$ where the final states are highly sensitive to the high-frequency modulations of $V(\mathbf{r})$ and thus to the accurate representation of its real-space variations;

– Another step, the inclusion of the full X-matrix compared to the single-site approximation, presented in Fig. S1(c), considerably improves the description of the relative intensity variations in the $h\nu$ interval between 400 and 500 eV (magenta arrow, for example) but does not notably affect the spectral intensity at higher energies. This observation can be understood from analysis of scattering amplitude $f_k(\theta)$, giving more insight into the scattering process. The calculations of $f_k(\theta)$ for Ag by Sébilleau et al.[8], reproduced in Fig. S2, demonstrate that for $E_k$ above ~400 eV it is strongly dominated by forward scattering. In practice, this means that for these energies the electrons scatter essentially along the rows of atoms, justifying the single-site approximation for the multiple scattering;

– Finally, at the last step of our computational refinement presented in the Fig. S1(d), we increased $l_{max}$ from 3 to 5. As expected, not only has this returned a vivid pattern of the MBFS-induced replica bands and excessive spectral broadening at low $E_k$ (such as where marked by magenta and yellow arrows, respectively) but also pushed these effects to yet higher $h\nu$ up to 700 eV (magenta arrow). Further increase of $l_{max}$ would inflate the computational time beyond presently realistic.

Although these successive refinements of the computational scheme do move towards a better description of the experiment, the achieved agreement with the experimental results can only be considered as qualitative. We conjecture that the remnant deviation may trace back to quite small sensitivity of the total energy to high-frequency components of the crystal potential. Therefore, the total-energy minimization used to generate the self-consistent potential in the DFT calculations may not ensure sufficient accuracy of its high-frequency components which critically affect the hybridization and thus non-FE effects in the final states at high energies. The accuracy of the final states used in the ARPES calculations can in principle be verified independently from the initial states by calculating the LEED spectra and their fitting to the experiment using the methodology previously developed for very low energies (see Refs. 9–11 and the references therein). In any case, including the subtle details of $V(\mathbf{r})$ within the FP approach and the use of sufficiently large $l_{max}$ give the best possible single-particle description of the photoemission final state.

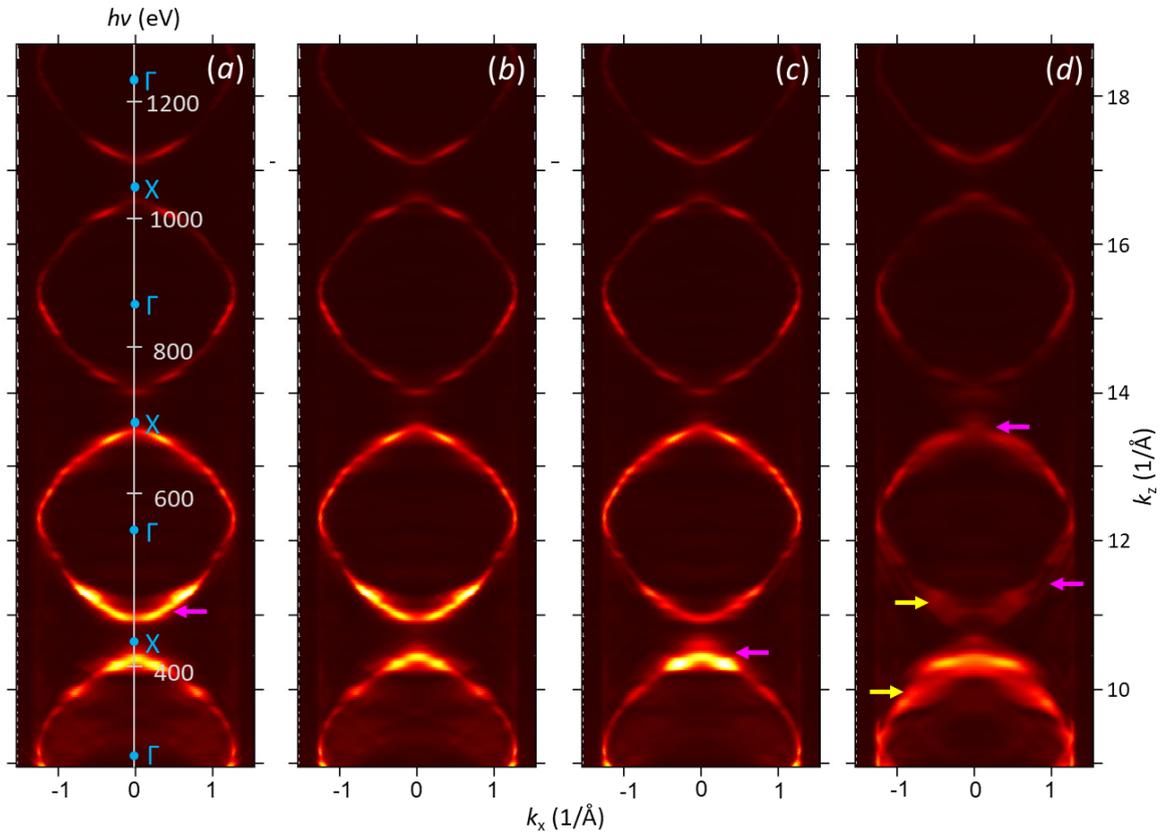

Fig. S1. One-step ARPES calculations as in Fig. 1(d) but using multiple-scattering final states under successive refinements of their treatment: (a) standard spherical-wave expansion and single-site scattering approximation; (b) adding full potential; (c) the full X-matrix beyond the single-site scattering; (d) increasing the angular momentum expansion to $l_{max}$ = 5. The calculations reproduce the multiple spectral peaks (magenta arrows) and the excessive spectral broadening (yellow) induced by the MBFSs.

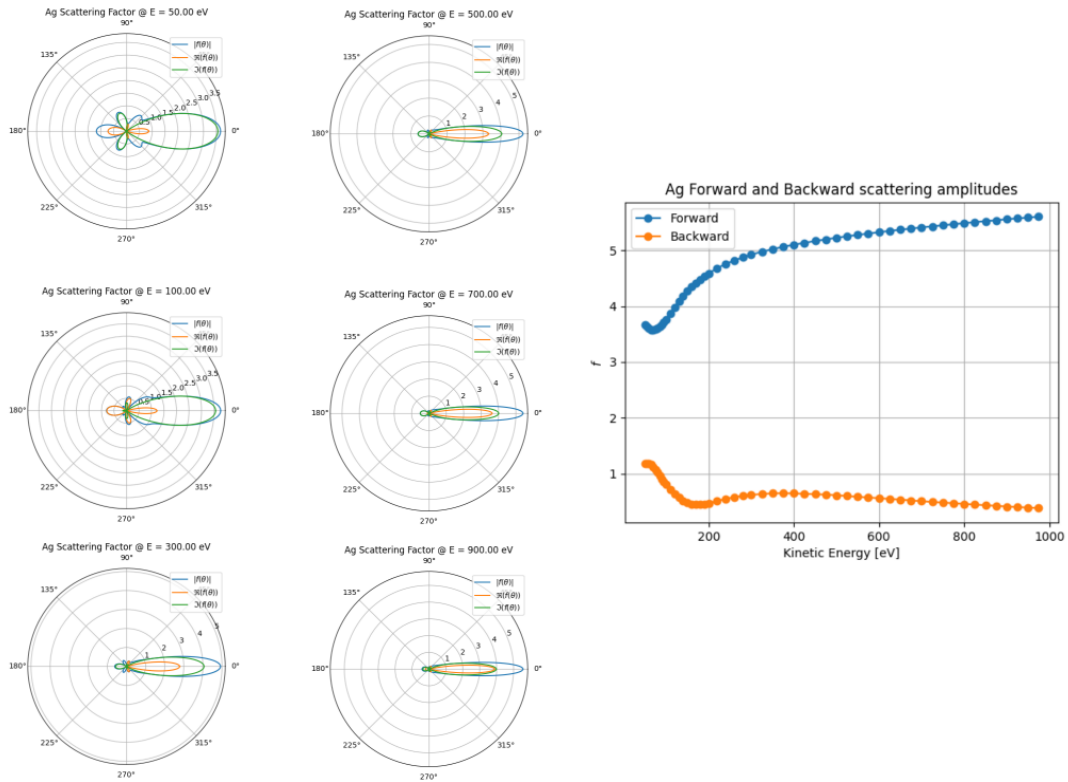

Fig. S2. Scattering amplitude $f_k(\theta)$ for Ag as a function of $E_k$ (left panels) and the total forward and backward scattering contributions (right panel).

# References

1. S. H. Vosko, L. Wilk, and M. Nusair, *Accurate Spin-Dependent Electron Liquid Correlation Energies for Local Spin Density Calculations: A Critical Analysis*, Canadian J. Phys. **58** (1980) 1200

2. H. Ebert, D. Ködderitzsch, and J. Minár, *Calculating Condensed Matter Properties Using the KKR-Green's Function Method – Recent Developments and Applications*, Rep. Prog. Phys. **74** (2011) 096501.

3. J. M. MacLaren, S. Crampin, D. D. Vvedensky, and J. B. Pendry, *Layer Korringa-Kohn-Rostoker Technique for Surface and Interface Electronic Properties*, Phys. Rev. B **40** (1989) 12164

4. J. Braun, J. Minár, and H. Ebert, *Correlation, Temperature and Disorder: Recent Developments in the One-Step Description of Angle-Resolved Photoemission*, Phys. Rep. **740** (2018) 1.

5. *Multiple Scattering Theory for Spectroscopies*, eds. D. Sébilleau, K. Hatada and H. Ebert. Springer Proc. Phys. **204** (2018).

6. J. Braun, *The theory of angle-resolved ultraviolet photoemission and its applications to ordered materials*. Rep. Prog. Phys. **59** (1996) 1267.

7. J. Braun, J. Minár, S. Mankovsky, V. N. Strocov, N. B. Brookes, L. Plucinski, C. M. Schneider, C. S. Fadley, and H. Ebert, *Exploring the XPS Limit in Soft and Hard X-Ray Angle-Resolved Photoemission Using a Temperature-Dependent One-Step Theory*, Phys. Rev. B **88** (2013) 205409

8. D. Sébilleau, S. Tricot, and A. Koide, unpublished (2022).

9. V. N. Strocov, H. Starnberg & P. O. Nilsson. *Excited-state bands of Cu determined by VLEED band fitting & their implications for photoemission*, Phys. Rev. B **56** (1997) 1717.

10. V. N. Strocov, R. Claessen, G. Nicolay, S Hüfner, A Kimura, A. Harasawa, S. Shin, A. Kakizaki, P.O. Nilsson, H.I. Starnberg & P. Blaha. *Three-dimensional band mapping by angle-dependent very-low-energy electron diffraction and photoemission: Methodology and application to Cu*. Phys. Rev. B **63** (2001) 20510.

11. V. N. Strocov, E.E. Krasovskii, W. Schattke, N. Barrett, H. Berger, D. Schrupp & R. Claessen. *Three-dimensional band structure of layered $TiTe_2$: Photoemission final-state effects*. Phys. Rev. B **74** (2006) 195125.